# Graphene Heat Spreaders and Interconnects for Advanced Electronic Applications

## Alexander A. Balandin

Nano-Device Laboratory (NDL) and Phonon Optimized Engineered Materials (POEM) Center, Department of Electrical and Computer Engineering, University of California – Riverside, Riverside, California 92521 USA
E-mail: balandin@ece.ucr.edu

**Invited Review**

Graphene revealed a number of unique properties beneficial for electronics, including exceptionally high electron mobility and widely tunable Fermi level. However, graphene does not have an electron energy band gap, which presents a serious hurdle for its applications in digital electronics. A possible route for practical use of graphene in electronics is utilization of its exceptionally high thermal conductivity and electron current conducting properties. This invited review outlines the thermal properties of graphene and describes prospective graphene technologies that are not affected by the absence of the energy band gap. Specific examples include heat spreaders, thermal coatings, high-current density electrodes and interconnects. Our results suggest that thermal management of advanced electronic devices can become the first industry-scale application of graphene.

## Introduction

Graphene [1] revealed a number of unique properties beneficial to electronics, including exceptionally high electron mobility and a widely tunable Fermi level [2]. However, graphene does not have an energy band-gap, which presents a serious hurdle for its applications in electronics. The efforts to induce a band-gap in graphene via quantum confinement or surface functionalization have not resulted in a major breakthrough. We have proposed several alternative applications that rely on electronic properties of graphene, but do not require the energy band gap. One example is non-Boolean logic gates implemented with "conventional" graphene transistors connected and biased in a configuration that provide negative differential resistance regions in the current-voltage characteristics [3]. Another example is selective "label-free" graphene sensors where the low-frequency current fluctuations are used as an additional sensing signal together with the channel resistance change [4]. However, these special niche applications will still require substantial time for research and development before they can get close to market introduction. In this invited review we outline several graphene technologies that rely on exceptional thermal [5] and current conducting properties of graphene [1-2], and which may require less time for market introduction. The absence of the electron energy band gap does not negatively affect the prospects of graphene heat spreaders, thermal coatings and high-current density conductors and interconnects. The described technological advancements suggest that the thermal management of electronics can become the first industry-scale application of graphene.

## Thermal Conductivity of Graphene and Few-Layer Graphene

In 2007, we discovered that the thermal conductivity of suspended single layer graphene can be exceptionally high [5-9]. The near room temperature (RT) values in a wide range from 2000 W/mK to 5000 W/mK were extracted under the assumption that the thickness of graphene is $h$=0.35 nm and that the heat transport in 20 μm length layers is diffusive or nearly diffusive [6]. It was established that the acoustic phonons make the dominant contribution to thermal conductivity of graphene. The value of thermal conductivity can change over orders of magnitude depending on the sample size, crystallinity, defect density and environment, e.g. suspended vs. supported, on a substrate or embedded in the matrix. The thermal conductivity of graphene has to be compared with that of the basal planes of bulk graphite, which is 2000 W/mK at RT for high-quality graphite [5]. The fact that the intrinsic thermal conductivity of graphene can be higher than the in-plane conductivity was explained by quenching of the phonon scattering processes in two-dimensional systems and resulting anomalously long mean free path of the low-frequency acoustic phonons in graphene [5, 10-12]. For practical thermal applications, few-layer graphene (FLG) can have certain benefits as compared to single layer graphene. In the thermal context, we consider a flake to be FLG rather than a piece of graphite as long as its thickness is below 7-10 atomic planes, and correspondingly, Raman spectrum is different from that of bulk graphite. The thermal conductivity of FLG is still rather high (~1000 W/mK – 2000 W/mK) and is subject to less degradation when FLG flake is imbedded inside matrix material or placed on a substrate as compare to that of graphene [13-16]. The larger thickness of FLG translates to higher heat fluxes. The excitement generated by graphene's properties led to a major progress in graphene and FLG synthesis using chemical vapor deposition (CVD), liquid phase exfoliation (LPE), metal-carbon melts and other techniques [17-21]. This progress, in its turn, created conditions for practical thermal applications of graphene.

## Graphene Laminate Thermal Coatings and Heat Spreaders

One of the graphene-based materials with the potential for near-term thermal applications is graphene laminate. Graphene laminate is made of the chemically derived graphene and FLG flakes, which are closely packed in overlapping structure. Graphene laminate can be deposited or "sprayed on" various surfaces and roll compressed. Potential applications include semiconductors packaging, back-end processing, thermal coatings for plastics used in solid-state lighting, and other systems where the low thermal conductivity of plastic presents a major hurdle. The physics of heat conduction in graphene laminate is complicated given the random nature of graphene flakes overlapping regions, a large distribution of the flake sizes and thicknesses as well as presence of defects and disorder. We investigated graphene laminate on polyethylene terephthalate (PET) substrates [22]. It was found that the thermal conductivity varies in the range from 40 W/mK to 90

W/mK at RT. The average size and the alignment of the graphene flakes are more important parameters defining the heat conduction than the mass density of the graphene laminate. The thermal conductivity scales up linearly with the average graphene flake size in both as deposited and compressed laminates. The compressed laminates have higher thermal conductivity for the same average flake size owing to better flake alignment. The possibility of more than *two orders-of-magnitude* enhancement of the thermal conductivity of plastic materials by coating them with thin graphene laminate can be used for improving thermal management of electronic and optoelectronic packaging. Figure 1 shows the cross-sectional scanning electron microscopy (SEM) images of the graphene laminate on PET film and its thermal conductivity as a function of the average flake size. Theory suggests that increasing the size of the graphene flakes and improving alignment can increases the thermal conductivity of graphene laminate beyond that of conventional semiconductors.

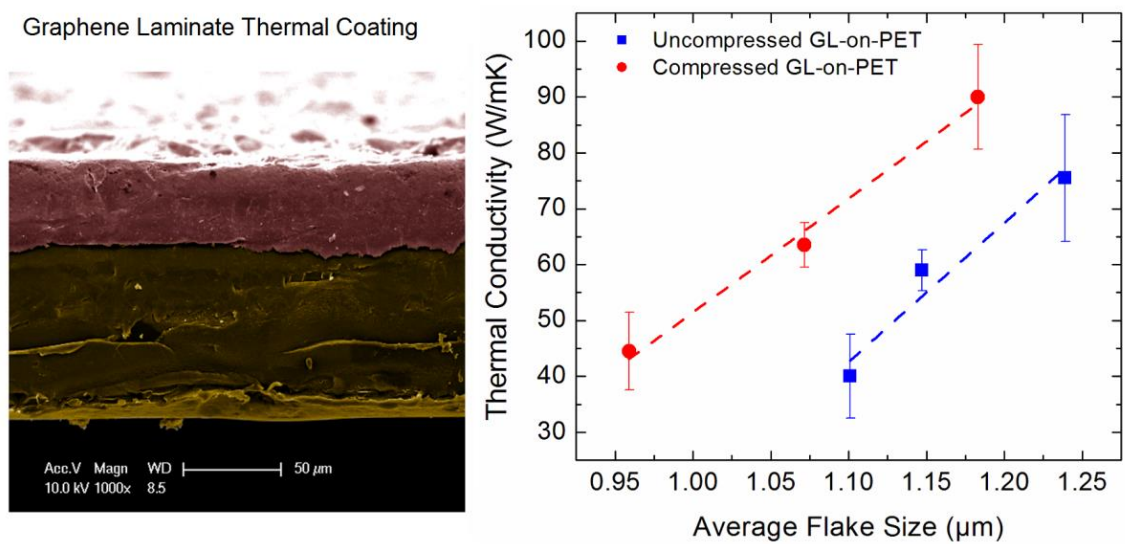

**Figure 1:** Cross-sectional SEM image of graphene laminate on PET film (left panel). The pseudo colors are used to indicate the graphene laminate (burgundy) and PET (yellow) layers. Thermal conductivity of graphene laminate as a function of the average flake size (right panel). The results are shown for the compressed (red circles) and uncompressed (blue rectangles) samples. For the same flake size, the compressed samples have higher thermal conductivity than uncompressed ones owing to better flake alignment. The data are after H. Malekpour, K.-H. Chang, J.-C. Chen, C.-Y. Lu, D. L. Nika, K. S. Novoselov and A. A. Balandin, *Nano Letters*, **14**, 5155 (2014).

### Few-Layer Graphene Interconnects on Synthetic Diamond

A number of research groups proposed graphene and FLG for transparent electrodes and interconnect applications capitalizing on graphene's current carrying ability [23-26]. Prototype graphene electrodes and interconnects built on $SiO_2/Si$ substrates reveal the breakdown current density of ~1 $\mu A/nm^2$, which is ~100× larger than the fundamental electromigration limit for the metals [27]. The breakdown mechanism in graphene is different from that in metals. We have demonstrated that by replacing $SiO_2$ with synthetic diamond one can increase the breakdown current density of FLG by *more than an order-* *of-magnitude* to ~18 µA/nm$^2$ (see Figure 2). Synthetic diamond improves heat conduction at high temperature, thus preventing the thermally induced breakdown. In synthetic ultra-nano-crystalline diamond, the thermal conductivity grows with temperature owing to increasing inter-grain transparency for the acoustic phonons that carry heat [27]. As a result the thermally-activated breakdown, which happens at high temperature, is shifted to much larger electrical current densities. Overall, synthetic diamond is a natural candidate for the use as a bottom dielectric in graphene devices, which can perform the additional function of an electrically insulating heat spreader. Recent years witness a major progress in CVD diamond growth performed at low temperature compatible with Si complementary metal-oxide-semiconductor (CMOS) technology. Direct thermal growth of sp$^2$ graphene from sp$^3$ synthetic diamond would allow for development of sp$^2$-on-sp$^3$ technology. Another possible related application of graphene and FLG is local heat spreader for GaN devices, which is used for decreasing the temperature of the hot spots [28]. One can envision device structures where graphene simultaneously plays a role of the interconnects and heat spreaders [29].

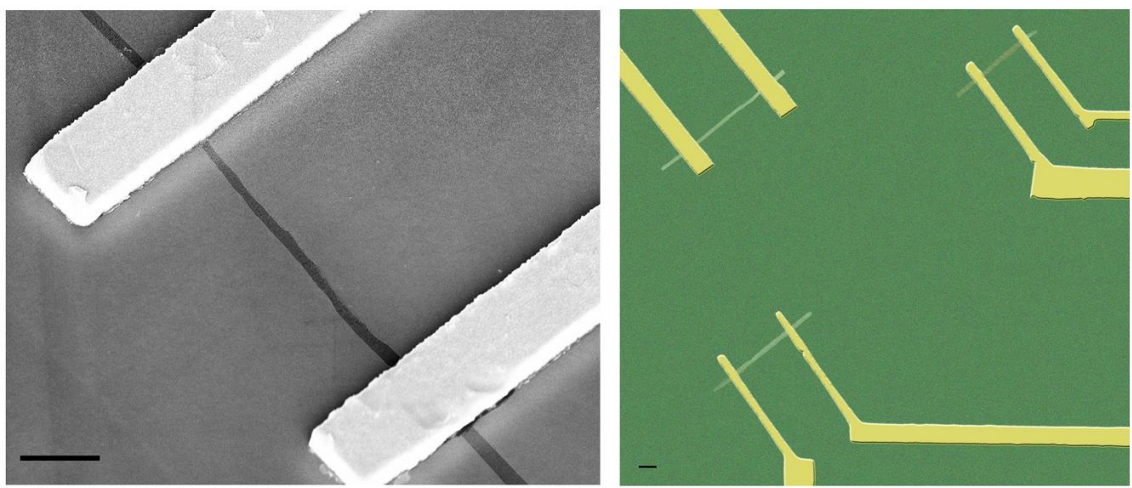

Figure 2: Scanning electron microscopy (left panel) and optical microscopy (right panel) images of the prototype interconnects on synthetic diamond. The two-terminal devices were used for the breakdown current density testing. The scale bar is 2 µm. The data are after J. Yu, G. Liu, A.V. Sumant, V. Goyal and A.A. Balandin, *Nano Letters*, **12**, 1603 (2012).

**Hybrid Graphene – Copper Interconnects and Heat Spreaders**

Copper became the crucial material for interconnects in Si CMOS technology by replacing Al. Main challenges with continuous downscaling of Si CMOS technology include electromigration in Cu interconnects, Cu diffusion to adjacent layers and heat dissipation in the interconnect hierarchies separated from a heat sink by many layers of dielectrics [30]. Combining graphene and Cu in some sort of hybrid heterogeneous global interconnect can bring potential benefits of reducing Cu electromigration and diffusion. Graphene capping of Cu interconnects increases the current density and reduces electrical resistance. Intersecting hybrid graphene – Cu interconnects have been shown to offer benefits for downscaled electronics [24-26]. Increasing the heat conduction properties of Cu films with graphene coating could become a crucial added benefit for improving the thermal management of the interconnect hierarchies. We demonstrated experimentally that graphene – Cu – graphene heterogeneous films reveal strongly enhanced thermal conductivity as compared to the reference Cu and annealed Cu films [31]. Chemical vapor deposition of a single atomic plane of graphene on both sides of Cu films increases their thermal conductivity by up to 24% near RT (see Figure 3). Interestingly, the observed improvement of thermal properties of graphene – Cu – graphene hetero-films results primarily from the changes in Cu morphology during CVD of graphene rather than from graphene's action as an additional heat conducting channel. Enhancement of thermal properties of graphene capped Cu films is important for thermal management of advanced electronic chips and proposed applications of graphene in the hybrid graphene – Cu interconnect hierarchies. Graphene and FLG have also shown promise in solving the thermal management problems with GaN technology [28, 32].

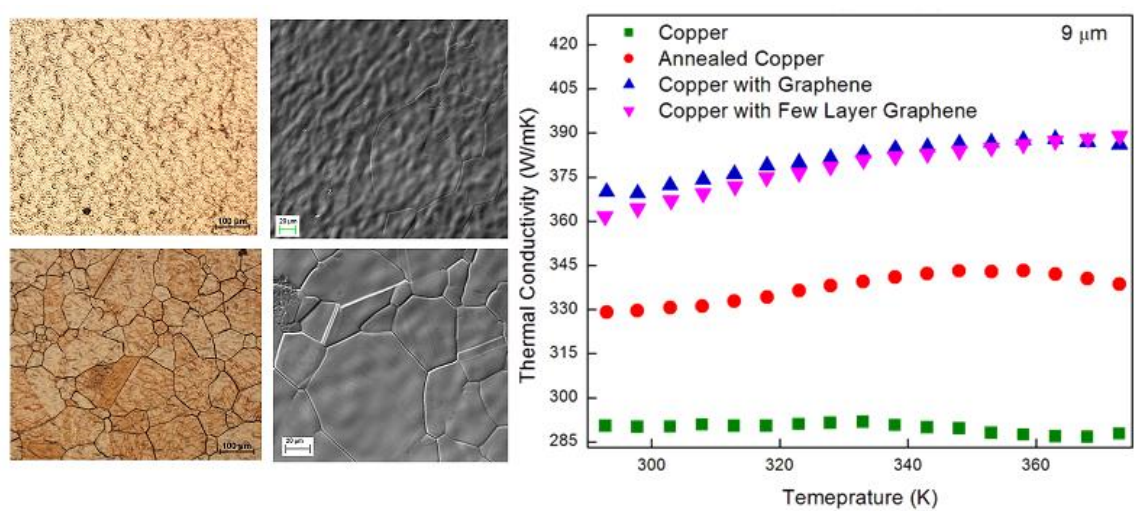

**Figure 3:** Scanning electron microscopy image of copper surface before CVD of graphene (top left panel) and after CVD of graphene (bottom left panel). Thermal conductivity of copper, annealed copper and copper after CVD of graphene (right panel). Note that CVD of graphene substantially increases the *apparent* thermal conductivity of graphene coated copper. The data are after P. Goli, H. Ning, X. Li, C.Y. Lu, K. S. Novoselov and A. A. Balandin, *Nano Letters*, **14**, 1497 (2014).

## Conclusions

We reviewed the thermal properties of graphene and described promising graphene technologies that are not affected by the absence of the energy band gap but rather utilize excellent heat conduction properties of graphene. The considered examples included heat spreaders, thermal coatings and high-current density interconnects. It is possible that the thermal management of advanced electronic devices can become the first industry-scale application of graphene.

*Acknowledgments*

The work at UC Riverside was supported, in part, by the National Science Foundation (NSF) project CMMI-1404967 Collaborative Research Genetic Algorithm Driven Hybrid Computational Experimental Engineering of Defects in Designer Materials; NSF project ECCS-1307671 Two-Dimensional Performance with Three-Dimensional Capacity: Engineering the Thermal Properties of Graphene, and by the STARnet Center for Function Accelerated nanoMaterial Engineering (FAME) – Semiconductor Research Corporation (SRC) program sponsored by The Microelectronics Advanced Research Corporation (MARCO) and the Defense Advanced Research Project Agency (DARPA).